\documentclass[12 pt]{article}

\usepackage[russian]{babel}
\usepackage{graphicx}

\AtBeginDocument{
  \pretolerance-1
  \tolerance1900
  \adjdemerits10000
  \emergencystretch10mm   
  \clubpenalty10000       
  \widowpenalty10000
  \displaywidowpenalty10000
  }

\author{A.\,P. Saiko\/\thanks{saiko@ifttp.bas-net.by}
        }

\title{\bf\normalsize KRYLOV-BOGOLIUBOV-MITROPOLSKY AVERAGING USED TO CONSTRUCT EFFECTIVE HAMILTONIANS IN THE THEORY OF STRONGLY CORRELATED ELECTRON SYSTEMS}
\date{}

\begin{document}
\maketitle
\begin{center}
Abstract
\end{center}
 \footnotesize We show that the
Krylov-Bogoliubov-Mitropolsky averaging in the canonical formulation
can be used as a method for constructing effective Hamiltonians in
the theory of strongly correlated electron systems. As an example,
we consider the transition from the Hamiltonians of the Hubbard and
Anderson models to the respective Hamiltonians of the t-J and Kondo
models. This is a very general method, has several advantages over
other methods, and can be used to solve a wide range of problems in
the physics of correlated systems.

 {\footnotesize {\bf PACS}:
71.10.Fd, 71.27.+a, 75.30.Mb}

Model Hamiltonians  used to describe strongly correlated electron
systems with the electron potential energy much greater than its
kinetic energy can be significantly simplified by reducing them to
effective Hamiltonians in spin variables. Such a simplification can
be realized by the initial Hamiltonian renormal-izations aimed at
eliminating high-energy states and at passing to a subspace with
lower energies of the quantum states. Eliminating high-energy states
is justified in this case because the system properties (e.g.,
electrical conduction, magnetization, etc.) under the usual
laboratory conditions are determined by the ground state and the
low-energy excitations. As a rule, the effective Hamiltonians are
constructed using the canonical transformation method [1], [2],
which allows eliminating an "off-diagonal" small perturbation
operator, which is responsible for transitions between low- and
high-energy states, from the original Hamiltonian. Sometimes, the
perturbation theory is also used in systems with degenerate states
[3]. Here, we propose a method for constructing effective
Hamiltonians based on using the Krylov-Bogoliubov-Mitropolsky (KBM)
averaging method [4] developed for solving problems in the theory of
nonlinear oscillations. The averaging method allows eliminating the
rapidly oscillating terms (the high-energy states) in any
perturbation order in the Hamiltonian written in the interaction
representation, which leads to an effective Hamiltonian written in
the approximately ''diagonal''  form. In what follows, as examples,
we show how the KBM averaging method in its canonical formulation
can be used to obtain the t-J model Hamiltonian from the Hubbard
Hamiltonian and the Kondo model Hamiltonian from the Anderson model
Hamiltonian.

The paradigmatic model in the theory of strongly correlated electron
systems is the Hubbard model. The Hubbard Hamiltonian is written as
[1], [2]:
\begin{equation}
H=-t\sum_{<ij>,\sigma}(c_{i\sigma}^{+}c_{j\sigma}+H.c.)+U\sum_{i}n_{i\uparrow}n_{i\downarrow}\equiv
H_{t}+H_{U},
 \label{eq:1}
\end{equation}
where $H_{t}$  is the kinetic (band) term describing the motion over
the lattice sites, $H_{U}$ is the operator of the Coulomb energy of
repulsion of two electrons at the same site; $c^{+}_{i\sigma}$ is
the operator of creation of an electron with the spin $\sigma$ at
the site $i$, and $t$ is the matrix element of electron transition
from a given site to its neighbor. In the case of large Coulomb
energy $U$, the appearance of two electrons at the same site is
energetically unfavorable, and the original band splits into two
Hubbard subbands: the upper and the lower (with a gap between them)
corresponding to the one-electron and two-electron states. In the
case of a half-filled band (one electron at the site,
$n=\sum_{\sigma}n_{\sigma}=1$), the Mott transition occurs for $U
\sim t$, i.e., the dielectric ground state appears, and an indirect
exchange coupling of antiferromagnetic type is established between
the electrons at the site. The so-called t-J model is thus realized,
which describes the propagation of holes in the lower subband
against the background of interacting spins for $n<1$. If the t-J
model is derived from the Hubbard model, then the Coulomb term is
usually taken as the zeroth-order approximation, and the kinetic
term is considered a perturbation. In the interaction
representation, where  $H_{U}$ is taken as the zeroth-order
Hamiltonian, the kinetic term is divided into parts describing the
energetically more favorable processes (in the interior of the
Hubbard subbands) and the less favorable processes (between the
Hubbard subbands). Indeed, in this representation,
$$
c_{j\sigma}(\tau)=e^{iH_{U}\tau}c_{j\sigma}e^{-iH_{U}\tau}=c_{j\sigma}(1-n_{j\bar{\sigma}})+e^{-iU\tau}c_{j\sigma}n_{j\bar{\sigma}}
$$
where $\bar{\sigma}=-\sigma$ and the kinetic term
$H_{t}(\tau)=e^{iH_{U}\tau}H_{t}e^{-iH_{U}\tau}$ has the manifestly
multiparticle character:
$$H_{t}(\tau)=-t\sum_{<ij>,\sigma}[(1-n_{i\bar{\sigma}})c_{i\sigma}^{+}c_{j\sigma}
(1-n_{j\bar{\sigma}})+n_{i\bar{\sigma}}c_{i\sigma}^{+}c_{j\sigma}n_{j\bar{\sigma}}+H.c.]-$$
$$-t\sum_{<ij>,\sigma}[n_{i\bar{\sigma}}c_{i\sigma}^{+}c_{j\sigma}(1-n_{j\bar{\sigma}})
+n_{j\bar{\sigma}}c_{j\sigma}^{+}c_{i\sigma}(1-n_{i\bar{\sigma}})]e^{iU\tau}-$$
$$-t\sum_{<ij>,\sigma}
[(1-n_{i\bar{\sigma}})c_{i\sigma}^{+}c_{j\sigma}n_{j\bar{\sigma}}+(1-n_{j\bar{\sigma}})c_{j\sigma}^{+}c_{i\sigma}n_{i\bar{\sigma}}]e^{-iU\tau}\equiv$$
\begin{equation}
\equiv H^{0}_{t}+H^{+}_{t}(\tau)+H^{-}_{t}(\tau). \label{eq:2}
\end{equation}
The first and second terms in  $H^{0}_{t}$ are responsible for the
electron kinetics in the respective lower and upper Hubbard
subbands;
 $H^{+}_{t}(\tau)$ describes the appearance of the second electron at a site already containing an
electron, i.e., the transition from a low-energy state to a
high-energy state (from the lower Hubbard subband into the upper);
and $H^{-}_{t}(\tau)$ represents the reverse process in which the
number of doubly occupied sites (sites with a pair of electrons)
decreases by unity. The existence of the rapidly oscillating factors
 $e^{ \pm iU\tau}$ in $H^{+}_{t}(\tau)$ and $H^{-}_{t}(\tau)$ additionally indicates that the processes with transitions between the Hubbard subbands are unfavorable.

The KBM averaging method allows eliminating the rapidly oscillating
''off-diagonal''  ~terms  $H^{\pm}_{t}(\tau)$ in any perturbation
order in $t$ (more precisely, in  $t/U$). and can be applied to
Hamiltonian (2). We briefly describe the method in the canonical
formalism [4]-[6].

We consider the Liouville equation for the density matrix of a
quantum system

\begin{equation}
\frac{\partial \rho}{\partial \tau}=-i[H_0+H_1,
\rho]\equiv-i(L_0+L_1)\rho,
 \label{eq:3}
\end{equation}
where $H_0$ is the unperturbed ''diagonal'' ~Hamiltonian of the
system, $H_1$ is a small perturbation, i.e., the ''off-diagonal
term,''~ and $L_0$ and $L_1$ are the Liouvillians corresponding to
the Hamiltonians. It follows from the condition $\|H_0\|\gg\|H_1\|$
where $\|\ldots\|$ denotes the value of an operator in frequency
units, that the fast motion with the period $2\pi/\|H_1\|$ is
superimposed on a slower process characterized by time of the order
of $\sim\|H_1\|^{-1}$. ¬ (\ref{eq:3}) We can pass to the interaction
representation  $\sigma =e^{iL_0t}\rho $,
$L_1(\tau)=e^{iL_0\tau}L_1e^{-iL_0\tau}$ in Eq. (3) and then apply
the KBM averaging method to the equation
\begin{equation}
\frac{\partial \sigma}{\partial \tau}=-iL_1(\tau)\sigma
 \label{eq:4}
\end{equation}
in order to eliminate the rapidly oscillating terms and to construct
an approximately ''diagonal''~ effective Hamiltonian (Liouvillian).
We briefly describe this procedure in the simplest form.

We note that because the function  $L_1(\tau)$, is periodic, it can
be expanded in the Fourier series
\begin{equation}
L_1(\tau)=\sum_nL_1^{(n)}e^{i\omega_n
\tau},\;\;\;\;\;\;L_1^{(n)}=\frac{1}{T}\int_0^Td\tau
L_1(\tau)e^{-i\omega_n\tau},\label{eq:5}
\end{equation}
where $\omega_n=2\pi n/T$ and $T$ is the period, which, in
particular, can coincide with the period $2\pi/\|H_0\|$ or be
multiple of it. Because we consider not the fast time ''vibration''
of the density matrix but its slow evolution, i.e., the motion
averaged over several time periods $T$, it is natural to define the
averaging operation
\begin{equation}
P^{\tau}\sigma (\tau)=\frac{1}{T}\int_0^Td\tau\;
\sigma(\tau)\equiv\langle\sigma(\tau)\rangle,
 \label{eq:6}
\end{equation}
where the projection operator $P^{\tau}$ is the operator of
averaging the rapidly varying quantities. We also define
$Q^{\tau}=1-P^{\tau}$ (it is easy to see that
$P^{\tau}P^{\tau}=P^{\tau}$, $Q^{\tau}Q^{\tau}=Q^{\tau}$, and
$P^{\tau}Q^{\tau}=Q^{\tau}P^{\tau}=0$), We can then write
\begin{equation}
\sigma(\tau)=P^{\tau}\sigma (\tau)+Q^{\tau}\sigma (\tau),
 \label{eq:7}
\end{equation}
i.e., we decompose the real motion of the system described by the
density matrix $\sigma(\tau)$ into the averaged  $P^{\tau}\sigma$
and fast ''vibration''~ $Q^{\tau}\sigma$. Substituting expression
(7) in (4) and acting on it from the left successively by the
operators $P^{\tau}$ and $Q^{\tau}$ we obtain the two coupled
equations
\begin{equation}
\frac{\partial}{\partial\tau}(P^{\tau}\sigma)=-iP^{\tau}L_1(\tau)P^{\tau}\sigma-iP^{\tau}L_1(\tau)Q^{\tau}\sigma,
 \label{eq:8}
\end{equation}
\begin{equation}
\frac{\partial}{\partial\tau}(Q^{\tau}\sigma)=-iQ^{\tau}L_1(\tau)Q^{\tau}\sigma-iQ^{\tau}L_1(\tau)P^{\tau}\sigma.
 \label{eq:9}
\end{equation}
Equations (8) and (9) were derived using the periodicity property of
the function $Q^{\tau}\sigma$ and the fact that $P^{\tau}\sigma$ is
a slow function of time. The solution of Eq. (9) can be represented
formally as
\begin{equation}
Q^{\tau}\sigma=-i\int^{\tau}d\tau'Q^{\tau'}L_1(\tau')P^{\tau'}\sigma(\tau')-i\int^{\tau}d\tau'Q^{\tau'}L_1(\tau')Q^{\tau'}\sigma(\tau'),
 \label{eq:10}
\end{equation}
and the operator constant in this expression is assumed to be zero:
we here use the freedom to choose this constant arbitrarily because
one first-order differential equation (4) was divided into two
equations, Eqs. (8) and (9), at the preceding stage. We iterate
expression (10) and obtain the power expansion in  $L_1$:
\begin{equation}
Q^{\tau}\sigma\approx\bigg\{-i\int^{\tau}d\tau'Q^{\tau'}L_1(\tau')-i\int^{\tau}d\tau'\int^{\tau'}d\tau''Q^{\tau'}L_1(\tau')Q^{\tau''}L_1(\tau'')\ldots,\bigg\}\langle\sigma\rangle,
 \label{eq:11}
\end{equation}
where the slowly varying function
$P^{\tau}\sigma\equiv\langle\sigma\rangle$ is brought outside the
integrand. Substituting formula (11) in (8), we obtain an equation
for slowly varying quantities. For example, in the second order in
$L_1$ , we have
\begin{equation}
\frac{\partial}{\partial\tau}\langle\sigma\rangle=-i\langle
L_1(\tau)\rangle\langle\sigma\rangle-\left\langle\int^{\tau}d\tau'\{L_1(\tau')(L_1(\tau')-\langle
L_1(\tau')\rangle)\}\right\rangle\langle\sigma\rangle\equiv-iL_{eff}\langle\sigma\rangle,
 \label{eq:12}
\end{equation}
where we introduce the effective Liouvillian
\begin{equation}
L_{eff}=\langle
L_1(\tau)\rangle-i\left\langle\int^{\tau}d\tau'\{L_1(\tau')(L_1(\tau')-\langle
L_1(\tau')\rangle)\}\right\rangle,
 \label{eq:13}
\end{equation}
which is associated with the effective Hamiltonian
\begin{equation}
H_{eff}=\langle
H_1(\tau)\rangle+\frac{i}{2}\left\langle\left[\int^{\tau}d\tau'(H_1(\tau')-\langle
H_1(\tau')\rangle),H_1(\tau)\right]\right\rangle.
 \label{eq:14}
\end{equation}
To derive expression (14) from (13), it is most convenient to use
Fourier expansion (5).

In the Hubbard model considered here, averaging the Hamiltonian
 $H_{t}(\tau)$ over the period $2\pi/U$ results in the relation
$\langle H_{t}(\tau)\rangle=H_{t}^{0}$ because the averages of
$H^{\pm}_{t}(\tau)$ are zero because of the factors  $e^{\pm
iU\tau}$. We use the total KBM averaging procedure, i.e., formulas
(14), to obtain
$$H^{(2)}_{eff}=H^{0}_{t}+\frac{1}{U}\langle[H^{+}_{t}(\tau),H^{-}_{t}(\tau)]\rangle+
\frac{1}{2U}\langle[H^{+}_{t}(\tau)-H^{-}_{t}(\tau),H^{0}_{t}]\rangle=
$$
\begin{equation}
=H^{0}_{t}+\frac{1}{U}[H^{+}_{t}(0),H^{-}_{t}(0)].
 \label{eq:15}
\end{equation}

We note that the canonical transformation method applied to the
Hubbard Hamiltonian leads to a similar effective Hamiltonian but
with additional ''off-diagonal''~ terms
$\sim[H^{\pm}_{t}(0),H^{0}_{t}]$ (see, e.g., [1], [2]), which can be
neglected because they take the interband transitions into account
only in the second order in
 $t/U$ [2] but not in the first order. In the KBM method, such terms do not appear, because the rapidly
oscillating term $[H^{+}_{t}(\tau)-H^{-}_{t}(\tau),H^{0}_{t}]$ in
(15) is zero after averaging over the period $2\pi/U$. Further, to
obtain the desired result, it is necessary to substitute
 $H^{0}_{t}$  given in (2) in expression (15) and to perform
rather simple transformations, which are described in detail in [1].
Namely, it is necessary to commute the operators, to omit the
three-site terms [1], [2], and to project the obtained Hamiltonian
on the lower Hubbard subband, i.e., to omit the terms multiplied on
the left and on the right by the operators $n_{i\bar{\sigma}}$ and
$n_{j\bar{\sigma}}$ such as, for example, the second term in $H^0_t$
(2): $n_{i\bar{\sigma}} c_{i\sigma}^+ c_{j\sigma} n_{j\bar{\sigma}}$
(the high-energy motion of electrons in the upper Hubbard subband is
described by similar terms). Ihe obtained effective Hamiltonian
$H^{(2)}_{eff}$ is just the Hamiltonian of the t-J model:

\begin{equation}
H^{(2)}_{eff}\rightarrow
H_{t-J}=-t\sum_{i,j,\sigma}[(1-n_{i\bar{\sigma}})c_{i\sigma}^{+}c_{j\sigma}(1-n_{j\bar{\sigma}})+h.c.]+J\sum_{ij}(\vec{S}_{i}\vec{S}_{j}-\frac{1}{4}n_{i}n_{j}),
\label{eq:16}
\end{equation}
where
$$\vec{S}_{i}=\frac{1}{2}\sum_{\sigma\sigma'}c_{i\sigma}^{+}\vec{\tau}_{\sigma\sigma'}c_{i\sigma'},$$
$\vec{\tau}$ is a vector composed of Pauli matrices,
$n_{i}=\sum_{\sigma}n_{i\sigma}$ and $J=4t^{2}/U$.

The algorithm described above for constructing an effective
Hamiltonian can also be used to transform the Hamiltonian of the
Anderson model into the Hamiltonian of the Kondo model. Transitions
from the Anderson model to the Kondo model are usually performed
using the Schrieffer-Wolff transformation [7], [8], i.e., the
canonical transformation with an anti-Hermitian operator whose form
must be chosen.

The KBM averaging method also has some obvious advantages in this
case. The original Anderson Hamiltonian can be written as [7], [8]

\begin{equation}
H=H_{0}+V,
 \label{eq:17}
\end{equation}
\begin{equation}
H_{0}=\sum_{k,\sigma}\varepsilon_{k}c_{k\sigma}^{+}c_{k\sigma}+\varepsilon\sum_{\sigma}
c_{d\sigma}^{+}c_{d\sigma}+Uc_{d\uparrow}^{+}c_{d\uparrow}c_{d\downarrow}^{+}c_{d\downarrow},
 \label{eq:18}
\end{equation}
\begin{equation}
V=\sum_{k,\sigma}\left(V_{kd}c_{k\sigma}^{+}c_{d\sigma}+h.c.\right),
 \label{eq:19}
\end{equation}
where $c_{k\sigma}^{+}$ and $\varepsilon_{k}$ are the creation
operator and the energy of an electron with the momentum $k$ and
spin  $\sigma$ in the conduction band, $c_{d\sigma}^{+}$ and
$\varepsilon$ are the creation operator and the energy of a
localized electron of the impurity atom, and $U$ is the Coulomb
interaction energy between two electrons occupying the impurity
atom. Here, $V$  describes the coupling that mixes the impurity
states with band states, and $V_{kd}$ is the coupling constant.

To pass to the interaction representation
$V(\tau)=e^{iH_{0}\tau}Ve^{-iH_{0}\tau}$, we first write
$c_{k\sigma}(\tau)=e^{iH_{0}\tau}c_{k\sigma}e^{-iH_{0}\tau}$ and
$c_{d\sigma}(\tau)=e^{iH_{0}\tau}c_{d\sigma}e^{-iH_{0}\tau}$ as
\begin{equation}
c_{k\sigma}(\tau)=c_{k\sigma}e^{-i\varepsilon_{k}\tau},
\,\,\,\,\,\,\,\,c_{d\sigma}(\tau)=c_{d\sigma}(1-n_{d\bar{\sigma}})e^{-i\varepsilon
\tau}+c_{d\sigma}n_{d\bar{\sigma}}e^{-i(\varepsilon+U)\tau}.
 \label{eq:20}
\end{equation}

As a result, we obtain
\begin{equation}
V(\tau)=\sum_{k,\sigma}\left\{V_{kd}[e^{i(\varepsilon_{k}-\varepsilon)\tau}c_{k\sigma}^{+}c_{d\sigma}(1-n_{d\bar{\sigma}})+e^{i(\varepsilon_{k}-\varepsilon-U)\tau}c_{k\sigma}^{+}c_{d\sigma}n_{d\bar{\sigma}}]+H.c.\right\}.
 \label{eq:21}
\end{equation}

We take all energies relative to the Fermi level. Then, in the
temperature region under study, the energies $\varepsilon_{k}$ (or
$\varepsilon_{k}-\mu$ instead of $\varepsilon_{k}$, where  $\mu$ is
the Fermi energy) are small, $\varepsilon$  is negative, and $U$,
$|\varepsilon |\gg\varepsilon_{k}$.

The time average of the Hamiltonian  $V(\tau)$ is zero in the
first-order of the perturbation theory because of fast oscillations
of the factors $e^{\pm i(\varepsilon_{k}-\varepsilon)\tau}$ and
$e^{\pm i(\varepsilon_{k}-\varepsilon-U)\tau}$. In the second order,
the KBM average of (14) is
$$H_{eff}^{(2)}(\tau)=\frac{i}{2}\left \langle
\left[\int^{\tau}d\tau'V(\tau'),V(\tau)\right]\right\rangle=
$$
$$
=-\frac{1}{2}\sum_{k,k',\sigma}[f(k,k',U)-f(k,k',0)](c_{k\sigma}^{+}c_{k'\bar{\sigma}}c_{d\bar{\sigma}}^{+}c_{d\sigma}-c_{k\sigma}^{+}c_{k'\sigma}
c_{d\bar{\sigma}}^{+}c_{d\bar{\sigma}})e^{i(\varepsilon_{k}-\varepsilon_{k'})\tau}+
$$

$$
+\frac{1}{2}\sum_{k,k',\sigma}f(k,k',0)c_{k\sigma}^{+}c_{k'\sigma}e^{i(\varepsilon_{k}-\varepsilon_{k'})\tau}-
$$
\begin{equation}
-\frac{1}{2}\sum_{k,\sigma}f(k,k,0)n_{d\sigma}-\frac{1}{2}\sum_{k,\sigma}[f(k,k,U)-f(k,k,0)]n_{d\sigma}n_{d\bar{\sigma}},
\label{eq:22}
\end{equation}
where
$$f(k,k',U)=V_{kd}V_{k'd}^{*}\left(\frac{1}{\varepsilon_{k}-\varepsilon-U}+\frac{1}{\varepsilon_{k'}-\varepsilon-U}\right),$$
and $e^{i(\varepsilon_{k}-\varepsilon_{k'})\tau}$ are slowly varying
factors. Passing to the Heisenberg representation in expression
(22), we obtain the effective Hamiltonian
\begin{equation}
H_{eff}^{(2)}(\tau)\rightarrow H_{eff,H}^{(2)}
=H_{0}+H_{eff}^{(2)}(0), \label{eq:23}
\end{equation}
where $H_{0}$ is defined by (18). If the Schrieffer-Wolff
transformation is used, then Hamiltonian (23) contains additional
terms of the form [7], [8]
\begin{equation}
H_{\delta}=-\frac{1}{2}\sum_{k,k',\sigma}\frac{V_{k'd}}{V_{k'd}^{*}}[f(k,k',U)-f(k,k',0)](c_{k\downarrow}c_{k'\uparrow}d_{\uparrow}^{+}d_{\downarrow}^{+}+h.c.),
\label{eq:24}
\end{equation}
which describe the high-energy processes, namely, the variation in
the impurity level population due to the capture of two band
electrons or the transition of two electrons from the d-orbital into
the conduction band. The effective Hamiltonian in the KBM method
does not contain such terms, because they are rapidly oscillating
terms in the interaction representation and hence vanish under the
averaging procedure. We neglect the terms describing the potential
scattering of the electron conduction (the second term) and the
electron energy renormalization on the impurity atom  (the third and
fourth terms) in  $H_{eff}^{(2)}(0)$ and also the Coulomb
(high-energy) term in $H_{0}$ and then obtain the Hamiltonian of the
Kondo model from (23) with expressions (18) and (22) taken into
account:
\begin{equation}
H_{eff,H}^{(2)}\rightarrow
H_{Kondo}=\sum_{k,\sigma}\varepsilon_{k}c_{k\sigma}^{+}c_{k\sigma}-\frac{1}{2}\sum_{k,k',\sigma}[f(k,k',U)-f(k,k',0)]
(c_{k\sigma}^{+}c_{k'\bar{\sigma}}c_{d\bar{\sigma}}^{+}c_{d\sigma}-c_{k\sigma}^{+}c_{k'\sigma}c_{d\bar{\sigma}}^{+}c_{d\bar{\sigma}}),
\label{eq:25}
\end{equation}
where the interaction term can be expressed in spin variables of the
band electrons and electrons on the d-orbital of the impurity atom
[7], [8]. The KBM averaging method can also be used to consider
systems under nonequilibrium conditions; for example, it can be used
to study the Kondo effect in the case where the energy of an
electron localized at a quantum dot is modeled by an external
alternating electric field. The KBM averaging is a reliable method
for constructing effective Hamiltonians for strongly correlated
electron systems: high-temperature superconductors, oxide magnetics
with colossal magnetoresistance, quantum dots, etc. We have
illustrated the use of this method with an example of the Hubbard
Hamiltonian transformation to the Hamiltonian of the t-J model and
the Anderson Hamiltonian transformation to the Hamiltonian of the
Kondo model. The computation algorithm is rather simple and natural.
This method does not require sophisticated tricks, which are needed
for choosing the specific form of the unitary operator if the
canonical transformation is used. It is also unnecessary to know the
eigenfunctions and eigenvalues of the Hamiltonian in the
zeroth-order approximation, in contrast to the case of the
perturbation theory used in systems with degenerate states. In
addition, the new method is self-sufficient: it does not result in
the appearance of "off-diagonal" terms of higher order in the
perturbation parameter in the transformed Hamiltonian, which
typically appear if the canonical transformation is used and are
then dropped based on some plausible reasoning. This approach is
very general and can be used to solve a wide range of problems in
the physics of strongly correlated electron systems. In particular,
it can be used to derive the effective Hamiltonian (of the t-J-model
type) for the recently discovered class of high-temperature
superconductors, which are iron-based layered compounds (see [9]).

REFERENCES

1.  P. Fazekas, Lecture Notes on Electron Correlation and
Magnetism, World Scientific, Singapore (1999).

2.  Yu. A. Izyumov, Phys. Uspekhi, 40, 521-523 (1997).

3.  C. L. Cleveland and R. Medina, Amer. J. Phys., 44, 44-46 (1976).

4.  N. N. Bogoliubov and Y. A. Mitropolsky, Asymptotic Methods in
the Theory of Non-linear Oscillations [in Russian], Nauka, Moscow
(1974); English transl. prev. ed., Hindustan Publishing Corp., Delhi
(1961).

5.  L. L. Buishvili and M. G. Menabde, Sov. Phys. JETP, 50,
1176-1180 (1979).

6.  A. P. Saiko, Phys. Solid State, 35, 20 (1993).

7.  J. R. Schrieffer and P. A. Wolff, Phys. Rev., 149, 491-492
(1966).

8.  P. Phillips, Advanced Solid State Physics, Westview, Boulder,
Colo. (2003).

9.  Yu. A. Izyumov and E. Z. Kurmaev, Phys. Uspekhi, 51, 1261-1286
(2008).

\end{document}